\documentclass[12pt]{article}
\usepackage[round]{natbib} 

\usepackage{graphicx,times}             
\usepackage{natbib}
\usepackage{amssymb,amsmath}
\bibpunct{(}{)}{;}{a}{}{,}

\usepackage[pagebackref=true]{hyperref}
\usepackage{hyperref}
\usepackage[section]{placeins}
\usepackage{threeparttable}
\usepackage{pdflscape}
\usepackage{rotating}
\usepackage{multirow}
\usepackage{booktabs}
\usepackage{graphicx}	
\usepackage{ulem} 
\usepackage{array,graphicx} 
\newcommand{\N}{$\textit{N}$}

\newcommand{\CO}{$^{12}$CO}
\newcommand{\HTWO}{$\textrm{H}_2$}
\newcommand{\CCO}{$^{13}$CO}

\newcommand{\one}{\uppercase\expandafter{\romannumeral1}}
\newcommand{\two}{\uppercase\expandafter{\romannumeral2}}
\newcommand{\three}{\uppercase\expandafter{\romannumeral3}}
\newcommand{\Jone}{$J=1 \rightarrow 0$}
\newcommand{\kms}{km\,s$^{-1}$}
\newcommand{\Jthree}{$J=3\rightarrow2$}
\newcommand{\tex}{$T_\textrm{ex}$}

\newcommand{\tk}{$T_\textrm{kin}$}

\newcommand{\abundance}{$\rm [CO/H_2]$}
\def \arcsec{\hbox{$^{\prime\prime}$}}

\def \sc{$^s$}

\newcommand\icarus{\ref@jnl{Icarus}}%
\newcommand\araa{{ARA\&A}}%
\newcommand\apj{{ApJ}}%
\newcommand\apjl{{ApJ}}%
\newcommand\apjs{{ApJS}}%
\newcommand\ao{\ref@jnl{Appl.~Opt.}}%
\newcommand\aap{{A\&A}}%
\newcommand\aapr{\ref@jnl{A\&A~Rev.}}%
\newcommand\aaps{{A\&AS}}%
\newcommand\azh{\ref@jnl{AZh}}%
\newcommand\baas{\ref@jnl{BAAS}}%
\newcommand\chjaa{\ref@jnl{ChJAA\,(Chin. J. Astron. Astrophys.)}}
\newcommand\cjaa {\ref@jnl{ChJAA\,(Chin. J. Astron. Astrophys.)}}
\newcommand\ibvs{\ref@jnl{IBVS}}                 
\newcommand\jrasc{\ref@jnl{JRASC}}%
\newcommand\memras{\ref@jnl{MmRAS}}%
\newcommand\mnras{{MNRAS}}%
\newcommand\pra{\ref@jnl{Phys.~Rev.~A}}%
\newcommand\prb{\ref@jnl{Phys.~Rev.~B}}%
\newcommand\prc{\ref@jnl{Phys.~Rev.~C}}%
\newcommand\prd{\ref@jnl{Phys.~Rev.~D}}%
\newcommand\pre{\ref@jnl{Phys.~Rev.~E}}%
\newcommand\prl{\ref@jnl{Phys.~Rev.~Lett.}}%
\newcommand\pasp{{PASP}}%
\newcommand\qjras{\ref@jnl{QJRAS}}%

\newcommand\skytel{\ref@jnl{S\&T}}%
\newcommand\solphys{\ref@jnl{Sol.~Phys.}}%
\newcommand\sovast{\ref@jnl{Soviet~Ast.}}%
\newcommand\ssr{\ref@jnl{Space~Sci.~Rev.}}%
\newcommand\zap{\ref@jnl{ZAp}}%
\newcommand\aplett{\ref@jnl{Astrophys.~Lett.}}%
\newcommand\apspr{\ref@jnl{Astrophys.~Space~Phys.~Res.}}%
\newcommand\bain{\ref@jnl{Bull.~Astron.~Inst.~Netherlands}}%
\newcommand\fcp{\ref@jnl{Fund.~Cosmic~Phys.}}%
\newcommand\gca{\ref@jnl{Geochim.~Cosmochim.~Acta}}%
\newcommand\grl{\ref@jnl{Geophys.~Res.~Lett.}}%
\newcommand\jcp{\ref@jnl{J.~Chem.~Phys.}}%
\newcommand\jgr{\ref@jnl{J.~Geophys.~Res.}}%
\newcommand\jqsrt{\ref@jnl{J.~Quant.~Spec.~Radiat.~Transf.}}%
\newcommand\memsai{\ref@jnl{Mem.~Soc.~Astron.~Italiana}}%
\newcommand\nphysa{\ref@jnl{Nucl.~Phys.~A}}%
\newcommand\physrep{\ref@jnl{Phys.~Rep.}}%
\newcommand\physscr{\ref@jnl{Phys.~Scr}}%
\newcommand\planss{\ref@jnl{Planet.~Space~Sci.}}%
\newcommand\procspie{\ref@jnl{Proc.~SPIE}}%
          
\newcommand\na{\ref@jnl{New Astron.}}
\newcommand\nar{\ref@jnl{New Astron. Rev.}}
\newcommand\actaa{\ref@jnl{Acta Astronomica}}
\newcommand\jcap{\ref@jnl{J. Cosmol. Astropart. Phys.}}
\newcommand\pasa{\ref@jnl{PASA}}%

\newcommand\caa{\ref@jnl{Chinese Astronomy and Astrophysics}}

\begin{document}
\baselineskip 16pt
\begin{center}
\textbf{\Large Updated Inventory of Carbon Monoxide in The Taurus Molecular Cloud} \\

\vspace{1.5cc}
{Yan Duan$^{1,2}$, Di Li$^{1,3,4}$, Laurent Pagani$^{5}$, Paul F. Goldsmith$^{6}$, 
Tao-Chung Ching$^{7}$, Chen Wang$^{1}$, and Jinjin Xie$^{8}$}\\

\vspace{0.3 cm}

{\small $^{1}$National Astronomical Observatories, Chinese Academy of Sciences, Beijing 100101, China, {\it dili@nao.cas.cn}\\ 
$^{2}$University of Chinese Academy of Sciences, Beijing 100049, China\\
$^{3}$Research Center for Intelligent Computing Platforms, Zhejiang Laboratory, Hangzhou 311100, China\\
$^{4}$NAOC-UKZN Computational Astrophysics Centre, University of KwaZulu-Natal, Durban 4000, South Africa\\
$^{5}$LERMA \& UMR\,8112 du CNRS, Observatoire de Paris, PSL University, Sorbonne Universit\'{e}s, CNRS, 75014 Paris, France\\
$^{6}$Jet Propulsion Laboratory, California Institute of Technology, 4800 Oak Grove Drive, Pasadena, CA 91109, USA\\
$^{7}$Jansky Fellow, National Radio Astronomy Observatory, 1003 Lopezville Road, Socorro, NM 87801, USA\\
$^{8}$Shanghai Astronomical Observatory, Chinese Academy of Sciences, 80 Nandan Road, Shanghai 200030, China}

\end{center}
\vspace{1cc}

\begin{abstract}
  \noindent  The most extensive survey of carbon monoxide (CO) gas in the Taurus molecular cloud relied on \CO\ and \CCO\ \Jone\ emission only, distinguishing the region where \CO\ is detected without \CCO\ (named mask 1 region) from the one where both are detected (mask 2 region). We have taken advantage of recent \CO\ \Jthree\ JCMT observations where they include mask 1 regions to estimate density, temperature, and N(CO) with a LVG model. This represents 1395 pixels out of $\sim$1.2 million in the mark 1 region. Compared to \citet{2010ApJ...721..686P} results, and assuming a \tk\ of 30 K, 
we find a higher volume density of molecular hydrogen of 3.3$\rm \times\ 10^3$ $\textrm{cm}^{-3}$, compared to their 250--700 $\textrm{cm}^{-3}$ and a CO column density of 5.7$\rm \times\ 10^{15}\ \textrm{cm}^{-2}$, about a quarter of their value. The differences are important and show the necessity to observe several CO transitions to better describe the intermediate region between the dense cloud and the diffuse atomic medium. 
Future observations to extend the \CO\ \Jthree\ mapping further away from the \CCO--detected region comprising mask 1 are needed to revisit our understanding of the diffuse portions of dark clouds.

\vspace{0.95cc}
\parbox{24cc}{\textbf{Key words:} submillimetre: ISM --- ISM: molecules --- ISM: clouds
}
\end{abstract}

\section{Introduction}
Understanding star formation is one of the fundamental challenges for astrophysics. Observations indicate that stars are born in molecular clouds, relatively cold, dense regions in the interstellar medium, which exist widely throughout the Milky Way and other galaxies \citep[e.g.,][]{1970ApJ...161L..43W, 1987ApJ...322..706D, 2001ApJ...547..792D, 2007prpl.conf...81B, 2012ARA&A..50..531K}. 
The Taurus molecular cloud complex is a famous low-mass star forming region about 140 pc away from us \citep{2009ApJ...698..242T}, which is close to us and has been widely studied with carbon monoxide and its isotopologues \citep[e.g.,][]{1987ApJS...63..645U, 1995ApJ...445L.161M,1996ApJ...465..815O,1998ApJ...502..296O} and other molecules \citep[e.g.,][]{1995ApJ...453..293L,2002ApJ...575..950O,2004ApJ...606..333T,2017ApJ...843...63F}. 
The proximity of Taurus allows us to accurately measure molecular gas properties, such as carbon monoxide excitation and depletion, the column density and volume density of molecular hydrogen, the relationship between gas and dust, and so on. 
The past carbon monoxide survey has provided systematic measurements of the total column density of CO, covering the largest area of the Taurus molecular cloud so far \citep{2008ApJ...680..428G, 2010ApJ...721..686P}.

\citet{2008ApJ...680..428G} published the Taurus \CO\ \Jone\ and \CCO \ \Jone\ survey covering 100 deg$^2$ using the Five College Radio Astronomy Observatory (FCRAO) 14 m telescope. The data have become a treasure trove for the studies of turbulence \citep{2012MNRAS.420.1562H}, cloud evolution \citep{2010ApJ...721..686P}, filament formation \citep{2013A&A...554A..55H}, and stellar feedback \citep{2015ApJS..219...20L}. \citet{2008ApJ...680..428G} defined different regions within the Taurus molecular cloud, which they called ``mask regions", according to which isotopologues of carbon monoxide were detected. 
Using the mask in which only \CO\ but no \CCO\ emission was detected, \citet{2008ApJ...680..428G} developed a large velocity gradient (LVG) method \citep{1983ApJS...51..203G} for analyzing the column density of \CO\ and \HTWO, assuming a kinetic temperature of 15 K and an optically thick \CO\ \Jone\ line. \citet{2010ApJ...721..686P} combined the FCRAO CO survey data with 2MASS extinction data to further investigate the relationship between CO and \HTWO\ (derived from dust extinction) in Taurus. For the mask 1 region where only \CO\ was detected but no \CCO, they used the RADEX program in its LVG mode to calculate the column density of CO. They compared N(CO) to N(\HTWO) to derive a varying $\left [ ^{12}\textrm{CO} / \textrm{H}_2 \right ]$ abundance ratio.

At sub-millimeter wavelengths, the James Clerk Maxwell Telescope (JCMT)'s observations of mid-J CO with a high spatial dynamic range provide more possibilities for the accurate excitation and dynamic measurement of the molecular gas in Taurus. \citet{2010MNRAS.405..759D} published CO \Jthree\ maps of B213-L1495 cloud and detection of 23 outflows there. 
\citet{2023ApJ...943..182D} detected a particular molecular bubble-outflow structure using JCMT \CO\ \Jthree\ observations of the Taurus B18 cloud.

In this paper, we focus on the region where \CO\ is detected but not \CCO\ in individual pixels (mask 1), covering 55\% of the area in Taurus where carbon monoxide was detected (mask 1 and mask 2 regions). 
Using \CO\ \Jone\ and \Jthree\ data, we can provide better CO column density measurements at the edges of Taurus B18, HCl2, and B213--L1495 clouds. 
We describe the observations in Section \ref{sec:obs}. We provide the measurements of N(CO) and the comparison with \citet{2010ApJ...721..686P} in Section \ref{sec:nco}. We discuss the CO-derived \N(\HTWO)\ in Section \ref{sec:nh2}. We summarize our results in  Section \ref{sec:Conclusions}.

\section{Observations and Data}\label{sec:obs}
\begin{figure*}
    \centering
	\includegraphics[width=0.8\columnwidth]{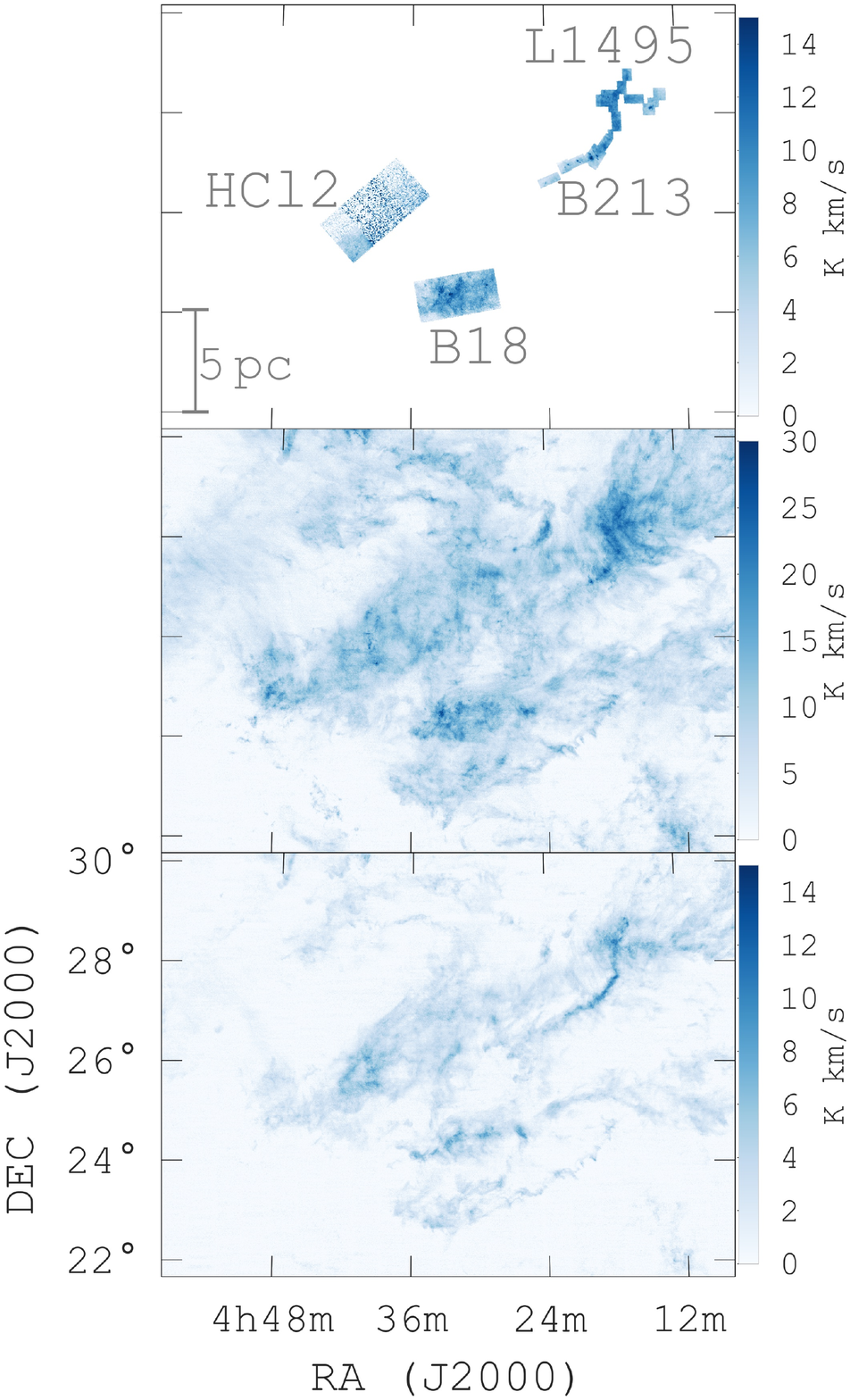}
    \caption{The \CO\ \Jthree\ (top), \CO\ \Jone\ (middle) and \CCO\ \Jone\ (bottom) \citep{2008ApJS..177..341N,2008ApJ...680..428G} data employed for analysis in the Taurus molecular cloud. The \CO\ \Jthree\ map of Taurus B213-L1495 cloud has been published by \citet{2010MNRAS.405..759D}.}
    \label{fig:total_data}
\end{figure*}
The data we used are displayed in Figure \ref{fig:total_data}. \CO\ \Jthree\ maps are all convolved with a Gaussian kernel to obtain an angular resolution of 45\arcsec, which is the Full Width at Half Maximum (FWHM) of the \CO\ \Jone\ map. We present CO \Jone\ and \Jthree\ observations in Sections \ref{sec:co10} and \ref{sec:co32}, respectively.

\subsection{\CO\ and \CCO\ \Jone\ from the FCRAO 14 m telescope}\label{sec:co10}
We utilize the \CO\ \Jone\ data observed with the 14 m Five College Radio Astronomy Observatory (FCRAO) millimetre telescope, which is extracted from the 100 deg$^2$ FCRAO large-scale survey covering the Taurus molecular cloud \citep{2008ApJS..177..341N,2008ApJ...680..428G}. 
The FWHM of the telescope beam is 45\arcsec\ for the \CO\ \Jone\  (115.271202 GHz) line and 47\arcsec\ for the \CCO\ \Jone\ (110.201353 GHz). 
The FCRAO 14 m telescope has a circular error beam of $\sim 0.5^{\circ}$ in diameter, contributing $\sim 25\%$ to the signal measured from a highly extended source much larger than the main beam \citep{2008ApJS..177..341N}.

\begin{figure}
    \centering
    \includegraphics[width=0.6\columnwidth]{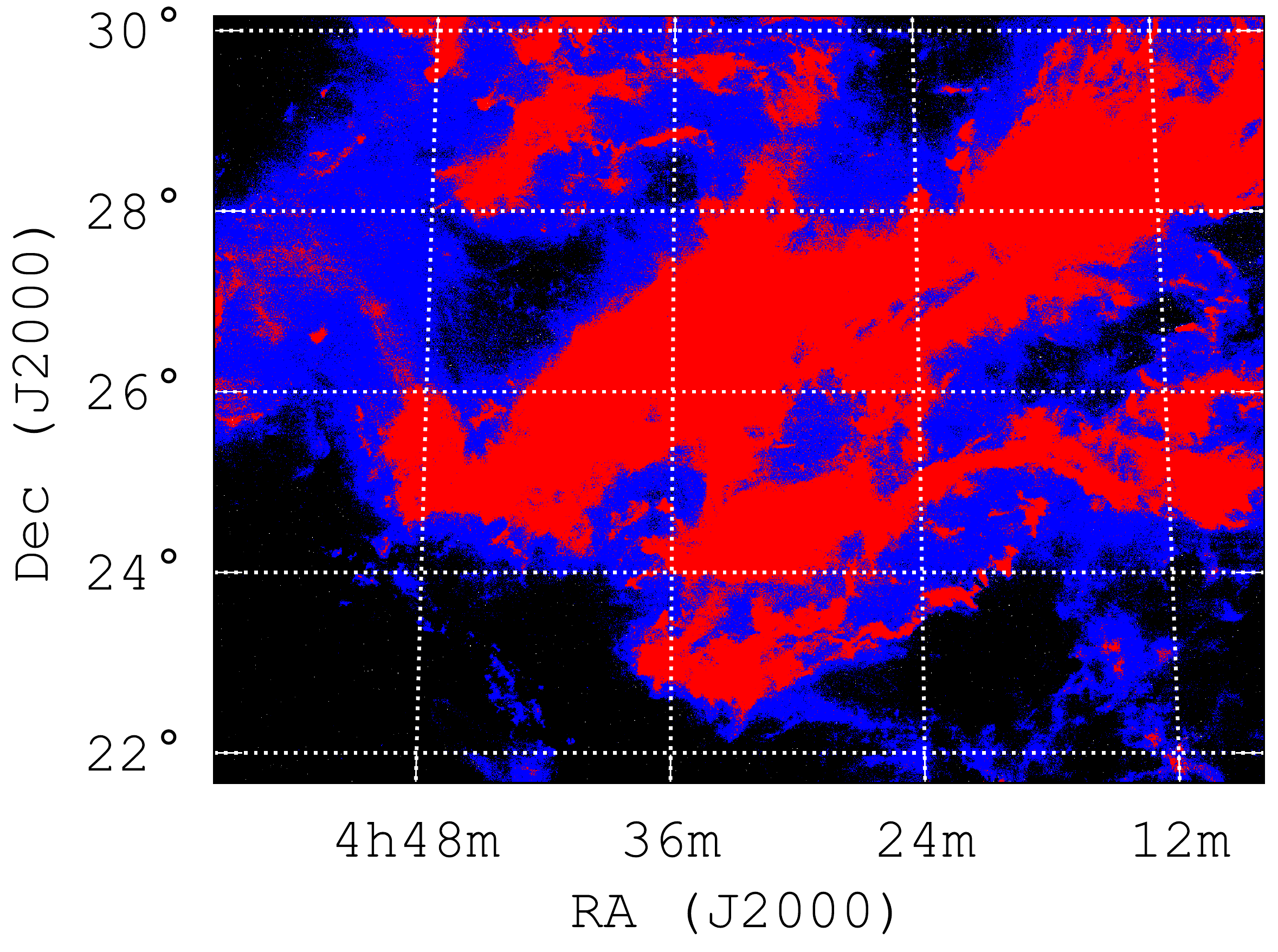}
    \caption{The mask regions in Taurus \citep{2008ApJ...680..428G,2010ApJ...721..686P}. Black represents mask 0, blue represents mask 1, and red represents mask 2.
    }
    \label{fig:mask}
\end{figure}

We follow the mask division of Taurus employed by \citet{2008ApJ...680..428G} and \citet{2010ApJ...721..686P}. The different mask regions are divided according to whether \CO\ and \CCO\ \Jone\ are detected or not \citep[their Figure 4 and Table 1]{2008ApJ...680..428G}. 
Mask 0 represents neither \CO\ nor \CCO\ detected, 
Mask 1 represents \CO\ but not \CCO\ detected,
Mask 2 represents both \CO\ and \CCO\ detected. 
As shown in Figure \ref{fig:mask}, Mask 1 accounts for 38\% of the total Taurus survey area. 
Here, we focus on mask 1, which includes regions in which \CO\ is detected but \CCO\ is not, with $T_\textrm{A}^*$ sensitivities of 0.28 K and 0.125 K in velocity resolutions of 0.26 \kms\ and 0.27 \kms\ for the \CO\ and \CCO\ spectra, respectively \citep{2008ApJ...680..428G}.

For the FCRAO 14 m telescope, the main beam efficiency $\eta_\textrm{mb}$ is 45 $\%$ and 48$\%$, at 115 GHz and 110 GHz respectively, as determined from measurements of Jupiter \citep{2010ApJ...721..686P}. The forward scattering and spillover efficiency
$\eta_\textrm{fss}$ \citep{1981ApJ...250..341K} is determined by observations of the Moon, $\eta_\textrm{fss}\approx \eta_\textrm{Moon}=0.70$ \citep{2010ApJ...721..686P}. Correcting for $\eta_\textrm{fss}$ provides a lower limit to the true radiation temperature for reasonably spatially extended structures \citep{1998ApJS..115..241H}. 
For observations of the Taurus molecular cloud, the source is larger than the main beam, but not uniform over the Moon size of $0.5^{\circ}$. In most of the region, the coupling efficiency is between $\eta_\textrm{mb}$ and $\eta_\textrm{fss}$. Here we define the coupling efficiency as $\eta_\textrm{coupling}=(\eta_\textrm{mb} +\eta_\textrm{fss})/2$ and the temperature corrected for coupling efficiency $T_\textrm{c}=T_\textrm{A}^*/ \eta_\textrm{coupling}$ for our CO data. 
Thus we get $\eta_\textrm{coupling}=0.575$ for \CO\ \Jone\ and $\eta_\textrm{coupling}=0.59$ for \CCO\ \Jone.

\subsection{\CO\ \Jthree\ from the JCMT Telescope}\label{sec:co32}
We have \CO\ \Jthree\ data for three regions of the Taurus molecular cloud, including B213-L1495, HCl2, and B18 from JCMT-HARP observations. 
The B18 cloud data are our own observations. We obtained 14 hours of JCMT-HARP observation time in band 3 on September 6, 11, 13, November 14, 2017 and August 10, 2018 (Program ID: M17BP027, M18BP072). 
The \CO\ \Jthree\ map covers 1.4 deg$\rm ^2$ in the B18 cloud. 
Data for L1495-B213 and HCl2 are the released archive data downloaded from the Canadian Astronomy Data Centre (CADC) \footnote{www.cadc-ccda.hia-iha.nrc-cnrc.gc.ca/en/search/}. 
\citet{2010MNRAS.405..759D} published CO \Jthree\ data of the B213-L1495 cloud and gave a detailed analysis of the detected outflows and dense cores, as part of the JCMT legacy survey of nearby star-forming regions in the Gould Belt \citep{2007PASP..119..855W}.

The \CO\ \Jthree\ transition has a rest frequency of 345.79599 GHz.
The telescope angular resolution is 14\arcsec\ at this frequency, corresponding to 0.0098 pc at a distance of 140 pc. 
The data have been processed with the Starlink package \citep{2014ASPC..485..391C}. For JCMT HARP, through the observations towards the Moon in 2007, we adopt $\eta_\textrm{mb}=61\%$ and $\eta_\textrm{fss}\approx \eta_\textrm{Moon}=77\%$ \citep{2009MNRAS.399.1026B}. We have $\eta_\textrm{coupling}=0.69$ for \CO\ \Jthree. The correction for $T_\textrm{A}^*$ is 1/$\eta_\textrm{coupling} \approx$ 1.45. 
For the \CO\ \Jthree\ data in these three clouds, we convolve with a Gaussian kernel to 45\arcsec, and re-grid to the angular and velocity resolutions of FCRAO \CO\ \Jone\ data. 
We compare the root-mean-square (RMS) noise and summarize all CO observations in Table \ref{table:rms}.

\begin{table}[]\small
\caption{Basic information on CO observations}
\label{table:rms}
\renewcommand{\arraystretch}{1.45}
\setlength{\tabcolsep}{0.6mm}{
\begin{tabular}{@{}llllllllll@{}}
\toprule
Emission                  & Source                  & Year                       & Area                 & Tau$^{\rm a}$       & Angular                   & $\eta_\textrm{mb}$      & $\eta_\textrm{fss}$   & $\eta_\textrm{coupling}$$^{\rm b}$  & RMS$^{\rm c}$       \\
                          &                         &                            & (deg$^2$)               &           & Resolution                &                       &                       &                       & (K)       \\ \midrule
\multirow{3}{*}{\CO\ \Jthree} & B18                     & 2017                       & 2                    & 0.05-0.2  & \multirow{3}{*}{14\arcsec} & \multirow{3}{*}{0.61} & \multirow{3}{*}{0.77} & \multirow{3}{*}{0.69} & 0.63      \\
                          & HCl2                    & 2015                       & 2.8                  & 0.03-0.36 &                           &                       &                       &                       & 1.66      \\
                          & B213-L1495              & 2007-2009                  & 11.8                 & 0.05-0.13 &                           &                       &                       &                       & 0.08–0.22 \\
\CO\ \Jone                  & \multirow{2}{*}{Taurus} & \multirow{2}{*}{2003-2005} & \multirow{2}{*}{100} &           & 45\arcsec                  & 0.45                  & 0.70                  & 0.575                 & 0.28      \\
\CCO\ \Jone                  &                         &                            &                      &           & 47\arcsec                  & 0.48                  & 0.70                  & 0.59                  & 0.125     \\ \bottomrule
\end{tabular}}
\footnotesize
\begin{flushleft}
	\textbf{Notes}.\\
	$^{\rm a}$The optical depth at 225 GHz, $\tau(225)$, represents the atmospheric opacity at the time of the observations. $\tau(225)$ can be converted to precipitable water vapor (PWV) using the equation $\tau(225)$= 0.04PWV+0.017. The values of $\tau(225)$ are from \href{https://www.cadc-ccda.hia-iha.nrc-cnrc.gc.ca/en/search/}{CADC}.\\ 
    $^{\rm b}$We define the coupling efficiency as $\eta_\textrm{coupling}=(\eta_\textrm{mb} +\eta_\textrm{fss})/2$, where $\eta_\textrm{fss}$ and $\eta_\textrm{mb}$ for JCMT at 345 GHz and for FCRAO 14 m at 115 GHz and 110 GHz are adopted from \citet{2009MNRAS.399.1026B} and \citet{2010ApJ...721..686P}, respectively.\\
    $^{\rm c}$RMS $T_\textrm{A}^*$ in K for JCMT data in B18 and HCl2 clouds were estimated at an angular resolution of 45\arcsec and a velocity resolution of 0.26 \kms. The sensitivity of B213-L1495 \CO\ \Jthree\ (at the 0.05 \kms resolution) comes from \citet{2010MNRAS.405..759D}. The sensitivities of \CO\ and \CCO\ \Jone\ are from \citet{2008ApJ...680..428G}.
    
\end{flushleft}
\end{table}

\begin{figure*}
   \centering
   \includegraphics[width=0.9\columnwidth]{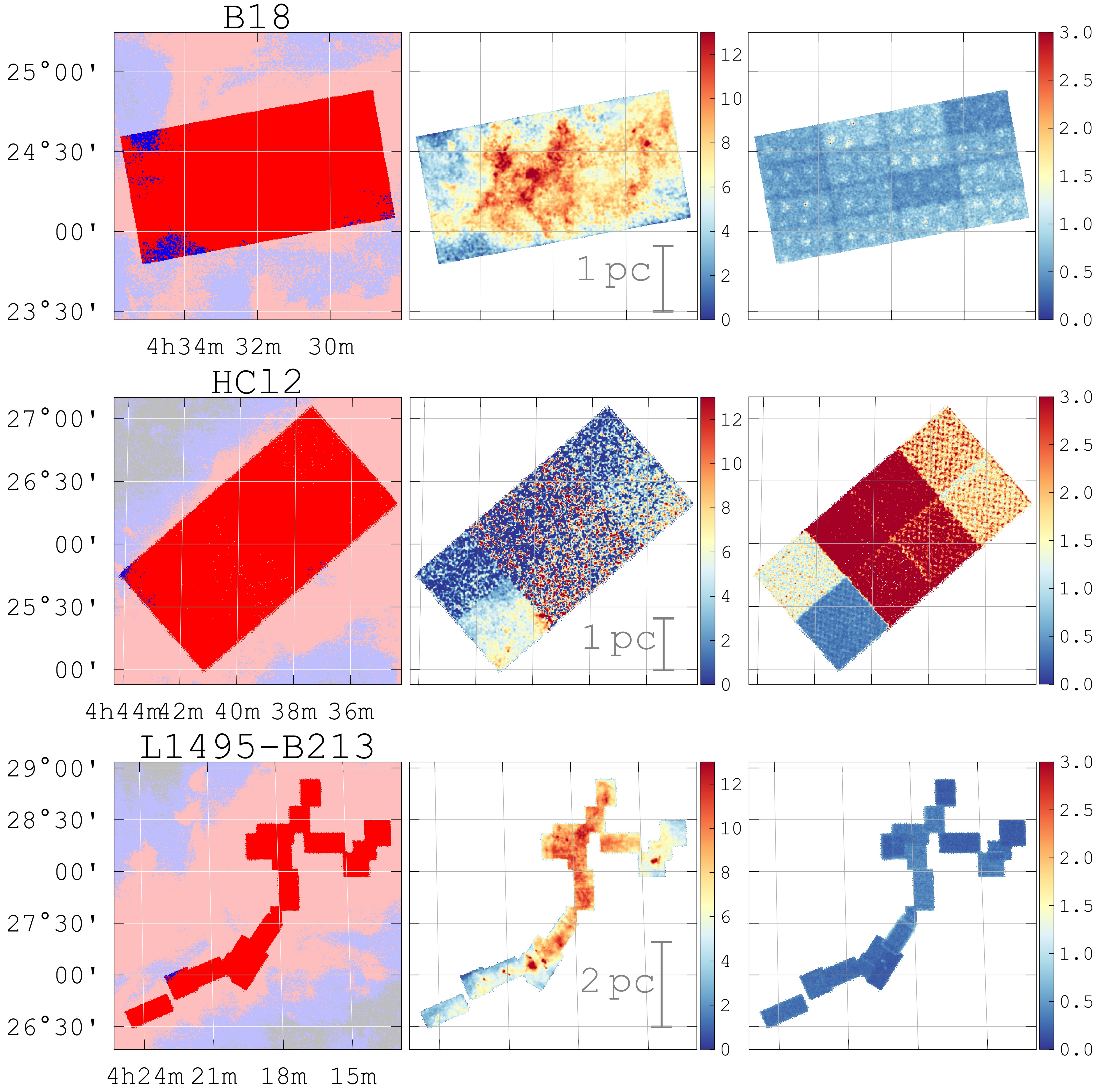}
   \caption{Images of \CO\ \Jthree\ of three clouds in Taurus. 
   The name of the cloud is labeled on the top of the first image in each row. 
    The left column represents the Mask maps. Mask 1 is defined by the blue-colored region.
    The middle column represents the integrated intensity map in the velocity range from 0 to 12 \kms. 
    The right column represents the sensitivity image of \CO\ \Jthree.}
    \label{fig:rms}
\end{figure*}

For \CO\ \Jone\ data, \citet{2008ApJ...680..428G} have identified the CO signal and divided the map into different masks. For \CO\ \Jthree\ data, we also performed the signal identification for the data in three clouds, shown in Figure \ref{fig:rms}. The main steps are as follows: (1) Draw an integrated intensity map in 0--12 \kms for each cloud. (2) Calculate the RMS noise for each pixel in the map. (3) Throw away the pixels for which the signal to noise is less than 3 $\sigma$ for each cloud. The remaining pixels are identified as having the \CO\ \Jthree\ signal. (4) Refer to the mask definition of \citet{2008ApJ...680..428G}, and select the \CO\ \Jthree\ pixels in mask 1. Using the above steps, we select a total of 1395 pixels for the three clouds. 

\section{\N(CO) for Mask 1}\label{sec:nco}
\citet{2010ApJ...721..686P} divided the Taurus molecular cloud into different masks to calculate \N(CO). 
For mask 2, where \CO\ and \CCO\ are detected in individual pixels and where density is high enough for the \Jone\ transitions to be thermalized, \N(\CO) can be determined from \CCO\ \Jone\ intensities under the local thermodynamic equilibrium (LTE) assumption. 
For mask 1, where only \CO\ \Jone\ is detected in individual pixels, LTE does not necessarily apply. \citet{2010ApJ...721..686P} binned the data into different excitation temperature ($T_\textrm{ex}$) intervals and calculated \N(CO) using the RADEX program under the optically thick ($\tau \gg 1$) \CO, a kinetic temperature of 15 K, and LVG assumptions.

Here we use \Jone\ and \Jthree\ transitions of \CO\ to independently estimate the \N(CO) of mask 1, with a LVG code (originally written by Dr. Jose Cernicharo) and similar to RADEX in its LVG mode but adapted to solve the \HTWO\ density and \CO\ column density from a pair of observed transitions \citep{1990A&A...234..469C}. There are only 1395 pixels with both \CO\ \Jone\ and \Jthree\ data that are located in the Taurus mask 1 region, at the edges of the B18, HCl2, and B213--L1495 clouds.

We reproduce the result of \citet{2010ApJ...721..686P} in mask 1, as shown in Section \ref{sec:radex}. Our LVG calculation with \CO\ \Jthree\ and \Jone\ data is given in Section \ref{sec:lvg}. We compare the results of the two studies in Section \ref{sec:3methods}.

\subsection{\N(CO) from \citet{2010ApJ...721..686P}}\label{sec:radex}
Instead of running the RADEX program, we only used the $N(\textrm{CO})/\delta v$ and \tex\ data in Table 1 of \citet{2010ApJ...721..686P} to restore the $N(\textrm{CO})/\delta v$. 
Here we calculate \tex\ using the maximum corrected  antenna temperature $T_\textrm{c}$ (peak) of \CO\ \Jone\ emission for every pixel according to 
\begin{equation}
    \begin{split}
    T_\textrm{ex}=\frac{5.53}{\textrm{ln}\left (1+\frac{5.53}{T_\textrm{c}(^{12}\textrm{CO})+0.83}\right )}
    \end{split}
    \label{tex}
\end{equation}
\citep[their Equation (19)]{2010ApJ...721..686P}. With this equation, we divide the \tex\ data of 1395 pixels into 8 different \tex\ bins. The width of each bin is 1 K. The median value in each bin ranges from 5.5 K to 12.5 K. For each \tex\ bin, all the data correspond to a value of CO column density per unit line width, $N(^{12}\textrm{CO})/\delta v$ (see Table 1 in \citealp{2010ApJ...721..686P}). $\delta v$ here is the observed FWHM of the line profile \citep{2010ApJ...721..686P}. We have recovered the \N(CO) from  $N(^{12}\textrm{CO})/\delta v$ and $\delta v$ of each pixel according to \citet{2010ApJ...721..686P}, and have summarized the median \N(CO) in Table \ref{table:nco}.

We define the uncertainty of \N(CO) calculated by the \citet{2010ApJ...721..686P} method from the RMS noise of the \CO\ \Jone\ temperature. We set $T_\textrm{c}$(\CO\ \Jone)$ + $RMS and $T_\textrm{c}$(\CO\ \Jone)$ - $RMS as the upper and lower limits of the data value range. The differences between the two derived values of \N(CO) and the original \N(CO) from $T_\textrm{c}$(\CO\ \Jone) are the upper and lower limits of the uncertainty.
\subsection{\N(CO) from LVG code with \CO\, \Jthree\, and \Jone}\label{sec:lvg}
Using the LVG statistical equilibrium method \citep[e.g.,][]{1960mes..book.....S,1974ApJ...189..441G}, our LVG code, expanded to include an inversion method \citep{1990A&A...234..469C}, has been modified to adopt the collisional rate coefficients from \citet{2010ApJ...718.1062Y}, the same as used by the RADEX program. These \CO\ collision rates include levels up to J=20 and \CO\ collisions with both para- and ortho-\HTWO. When we import the line width, the kinetic temperature \tk, and the temperatures of the two transitions for one molecule, this LVG code runs a grid of models. By inverting this grid, it returns $n$(\HTWO) and \N(CO) which are the best match for the two observed transitions.

Using this LVG code, we analyze the \CO\ excitation conditions through the observed $T_\textrm{c}$ with two \CO\ lines. We assume the \CO\ collisions with both para- and ortho-\HTWO\ molecules (assuming the ortho-to-para \HTWO\ ratio = 1), a \tk\ of 30 K, a background temperature of 2.7 K, and helium abundance of 0.1. 
One of the output parameters for LVG code is the column density per unit line width, $N(\textrm{CO})/\delta v$. 
We define the spatial variation of \CO\ line width $\delta v$ as $\left [  \int T_\textrm{c}(v)dv\right ]/T_\textrm{c}$(peak). Since there are two \CO\ spectral lines calculated in the code. We have two line widths, $\delta v$(\Jone) and $\delta v$(\Jthree). We take the arithmetic mean of the two. By measuring $\delta v$ in each pixel, we calculate the \N(CO) for all data pixel by pixel. 
For the B18, HCl2, and B213-L1495 regions in Taurus, there are a total of 1395 pixels in mask 1 for calculation. 
For the B213-L1495 cloud, \CO\ \Jthree\ emissions within mask 1 region are limited to the B213 cloud. The median \N(CO) for each cloud and the median \N(CO) in these three clouds are summarized in Table \ref{table:nco}.

\begin{figure}
    \centering
	\includegraphics[width=0.8\columnwidth]{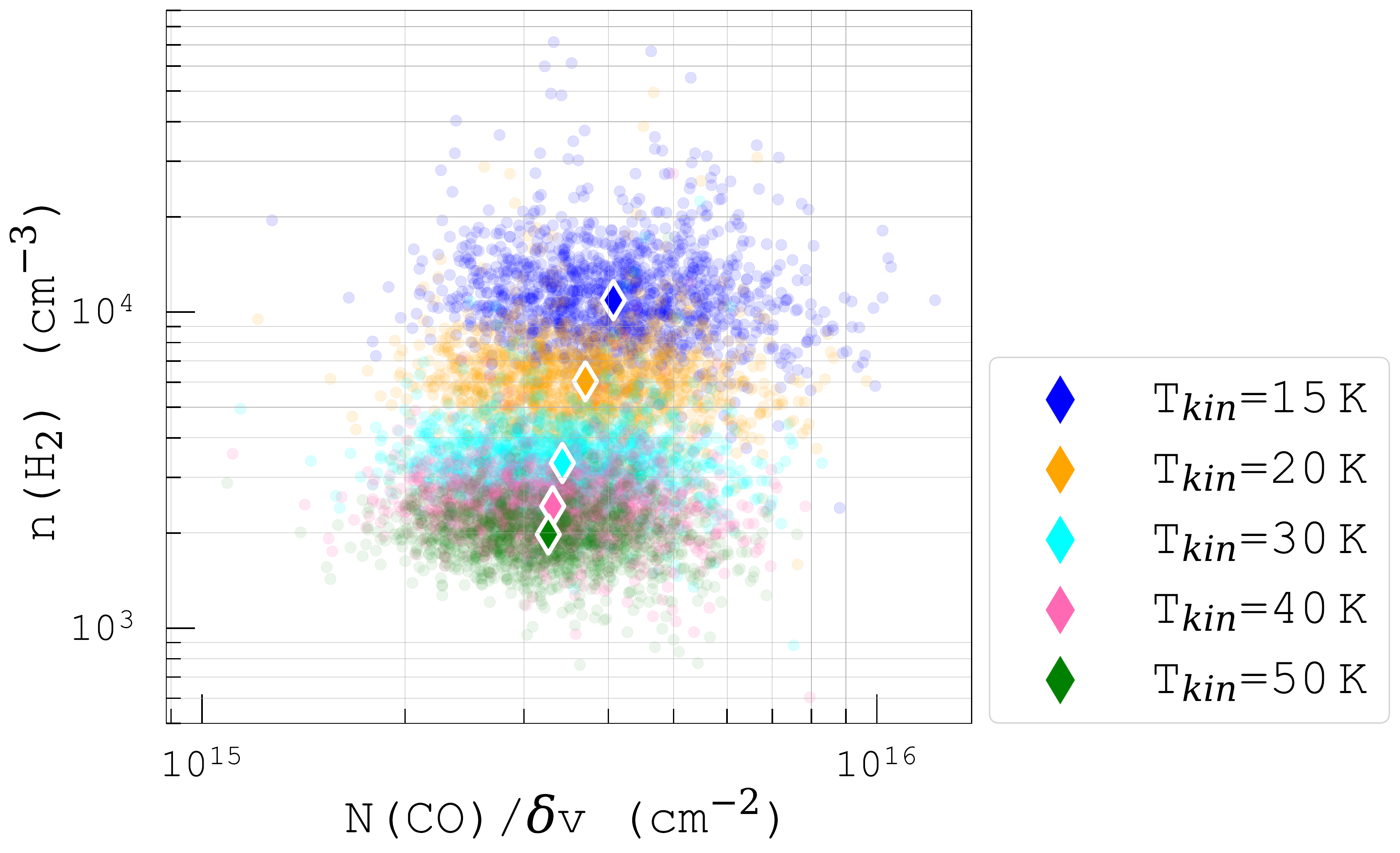}
    \caption{$n$(\HTWO) and $N(\textrm{CO})/\delta v$ for all data calculated under different \tk\ assumptions. The diamond of the same color represents the median of all data for each \tk.}
    \label{fig:tk}
\end{figure}

The kinetic temperature \tk\ cannot be measured directly at the edge of the cloud, 
because of a lack of data such as NH$_3$ hyperfine components \cite[e.g.][]{2020MNRAS.499.4432W,2021SCPMA..6479511X}. The LVG code requires us to assume a \tk. We find that the assumed \tk\ is anticorrelated  with the derived $n$(\HTWO). This trend has also been demonstrated in the past analysis of multi-level lines of CO with RADEX \citep{2013ApJ...774..134G}. We compare the results under different \tk\ assumptions, shown in Figure \ref{fig:tk}. 
The results indicate that when \tk$=$15 K, our LVG code calculates a large $n$(\HTWO) above 1$\rm \times10^{4}\ cm^{-3}$ because of the CO \Jthree\ data. When we assume \tk$\geqslant$30 K, $n$(\HTWO) drops below 3.3$\rm \times10^3\ cm^{-3}$ and is relatively close to each other, yielding more reasonable but still high densities. This may be an indication that the single density model adopted is not adequate. And more sophisticated modeling, including density inhomogeneities on a scale not resolved by telescope beams, is required but beyond the scope of the present study.
\N(CO)$/\delta v$ is about 4.1$\rm \times10^{15} - 3.3\times10^{15}\ cm^{-2}$ from 15 to 50 K, the difference of which is small. Here we assume \tk\ to be 30 K, the derived median $n$(\HTWO) is 3.3$\rm \times10^3$ $\textrm{cm}^{-3}$ and median $N(\textrm{CO})/\delta v$ is $3.4\times 10^{15}\ \textrm{cm}^{-2}$. When \tk\ is assumed to be 15 K as \citet{2010ApJ...721..686P}, $n$(\HTWO) would increase by 327\%, and $N(\textrm{CO})/\delta v$ would increase by 119\%. 
In dark cloud B5, \citet{1982ApJ...261..513Y} found that \tk\ rises from 15 K in the cloud center to 40 K at the cloud edge, with $n$(\HTWO) close to 2000 $\textrm{cm}^{-3}$ at the cloud edge, essentially the same as in the bulk of the cloud. Therefore, we consider that our assumption of 30 K for \tk\ in Mask 1 is reasonable.

\begin{figure}[htbp]
	\begin{minipage}[t]{0.49\linewidth}
		\centering
		\includegraphics[width=1\textwidth]{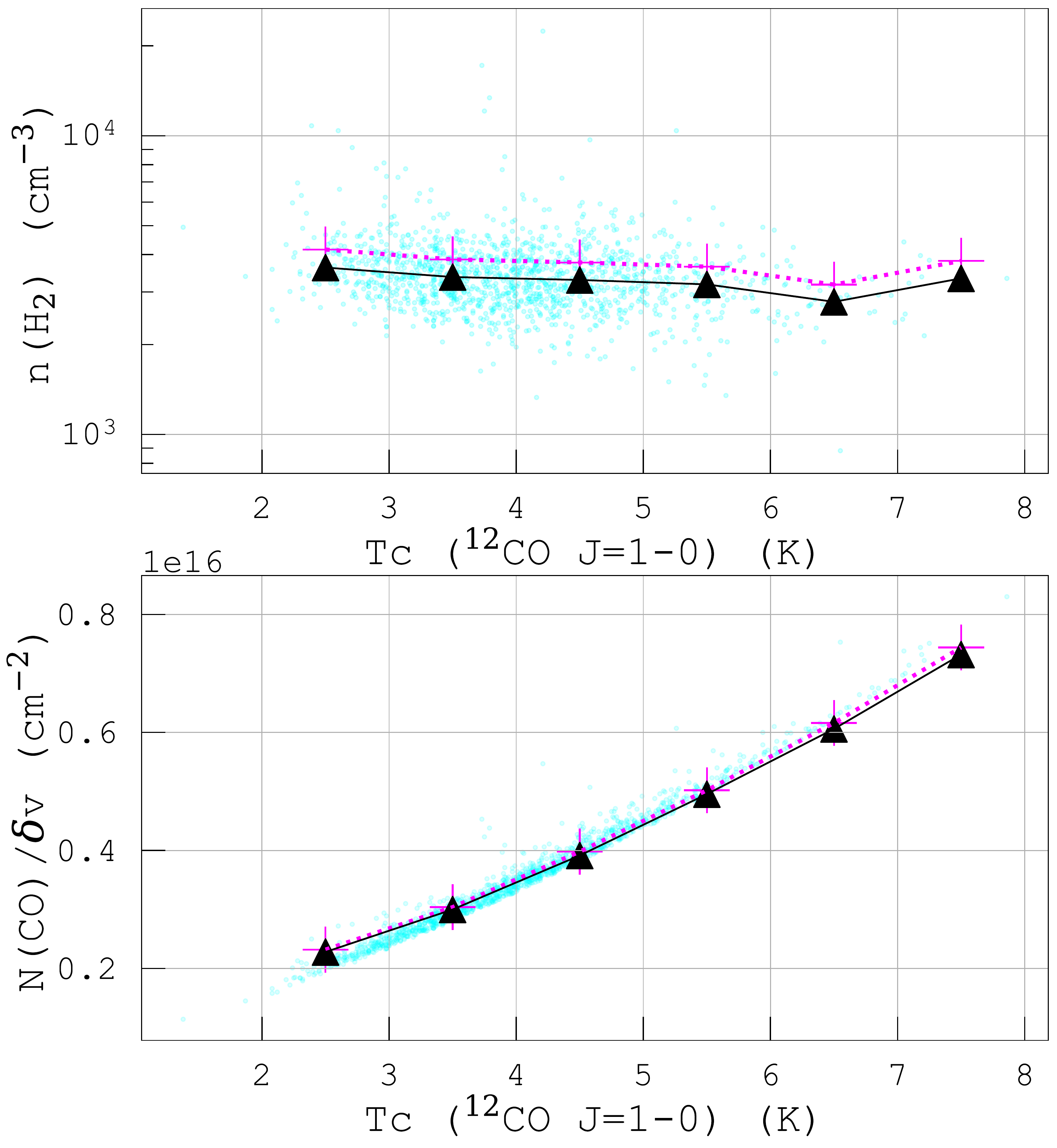}
		\caption{$n$(\HTWO) (top) and $N(\textrm{CO})/\delta v$ (bottom) of mask 1 binned by $T_\textrm{c}$ (\CO\ \Jone) (in 1 K bins). 
        The cyan points are the $T_\textrm{c}$, $n$(\HTWO), and $N(\textrm{CO})/\delta v$ determined for the 1395 pixels in mask 1.
        Black triangles and solid lines are the median of each $T_\textrm{c}$ bin under the assumption of ortho-to-para \HTWO\ ratio = 1. Magenta crosses and dotted lines are the median values under the assumption of ortho-to-para \HTWO\ ratio = 0.}
		\label{fig:lvg}
	\end{minipage}%
	\hfill
	\begin{minipage}[t]{0.49\linewidth}
		\centering
		\includegraphics[width=1\textwidth]{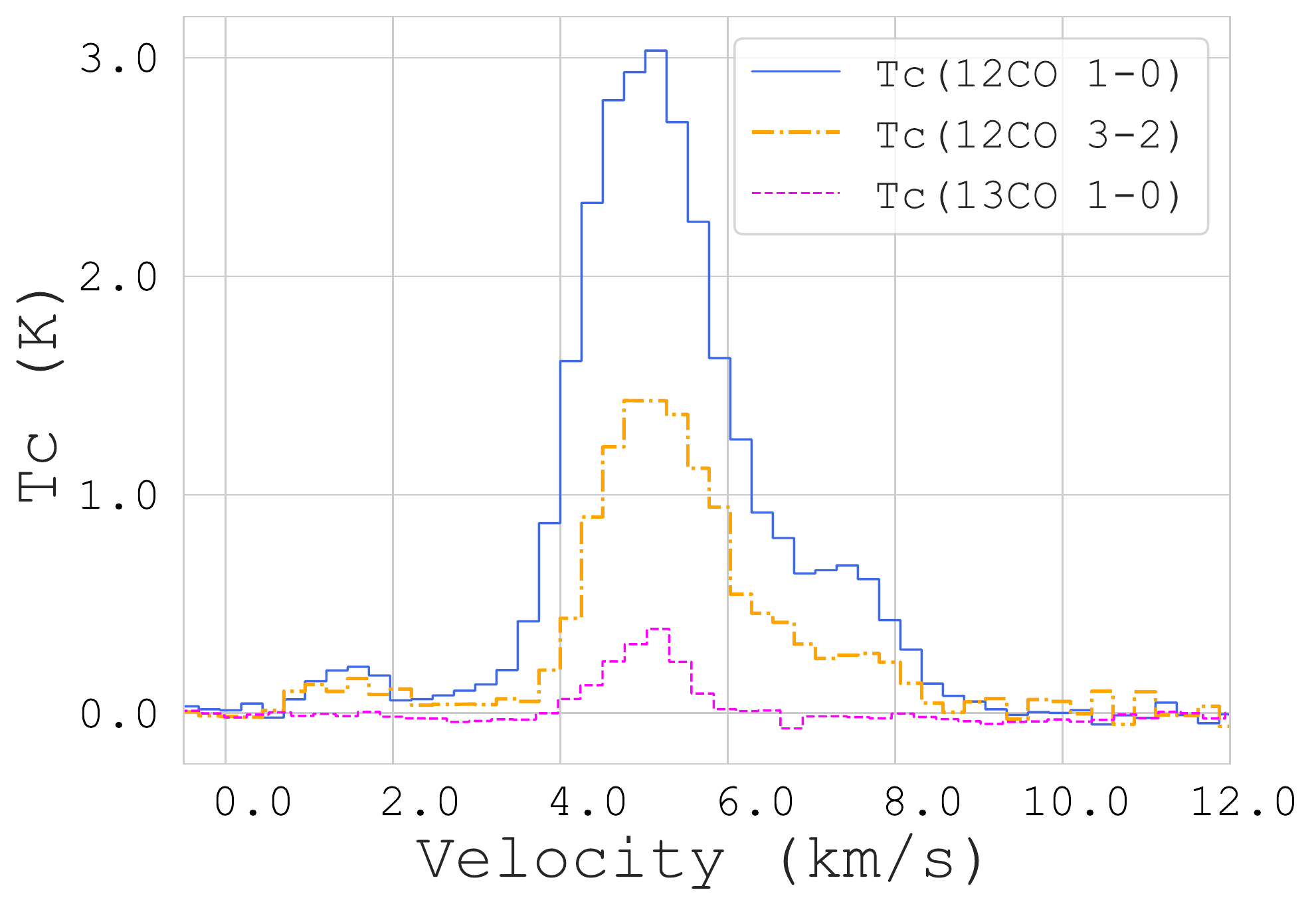}
		\caption{Average \CO, \CCO\ \Jone, and \CO\ \Jthree\ spectral lines for Taurus B18 cloud within mask 1 region, containing a total of 1276 pixels. }
    \label{fig:spec}
	\end{minipage}
\end{figure}

We bin $n$(\HTWO) and $N(\textrm{CO})/\delta v$ from LVG code into different $T_\textrm{c}$ (\CO\ \Jone) bins to show their trend with $T_\textrm{c}$, as shown in Figure \ref{fig:lvg}. 
For each $T_\textrm{c}$ bin, the median $n$(\HTWO) and $N(\textrm{CO})/\delta v$ are given and shown as black solid lines.
It is reasonable that $N(\textrm{CO})/\delta v$ increases steadily with the increase of $T_\textrm{c}$ (\CO\ \Jone), and the same trend is found in Figure 3 of \citet{2010ApJ...721..686P}. 
The value of $n$(\HTWO) does not change significantly with the increasing of $T_\textrm{c}$. It is largely determined by the observed ratio of the two \CO\ transitions, which is almost constant. 
The magenta dotted lines in Figure \ref{fig:lvg} represent the result under the assumption of 100\% para-\HTWO. Compared to collisions with 50\% para- and 50 \% ortho-\HTWO\ molecules (ortho-to-para \HTWO\ ratio = 1), when 100\% para-\HTWO\ is assumed (ortho-to-para \HTWO\ ratio = 0), $n$(\HTWO) becomes 114\% of its original value, and $N(\textrm{CO})/\delta v$ is practically constant.

We calculated the thermal pressure ($\left<P_\textrm{th}/k\right>=nT$) to be about $\rm 10^5\ K\ cm^{-3}$, which is almost the highest in the observed thermal pressure deduced from \CO\ and \CCO\ observations of molecular clouds in the Galactic plane, $\sim 10^4$ to $\rm 10^5\ K\ cm^{-3}$ \citep{1993AIPC..278..311S,2010ApJ...716.1191W}. 
This is possibly because the cloud here may be out of thermal equilibrium, so the pressure reflect approximate but incomplete thermal pressure balance. In Figure \ref{fig:spec}, the non-thermal line width of the average \CCO\ \Jone\ spectrum also demonstrate that the Taurus mask 1 region may deviate from the thermal pressure balance.

In our calculation, we mainly consider the uncertainties of \N(CO) and $n$(\HTWO) from three aspects, which have proportional or inverse effects on \N(CO) and $n$(\HTWO). Taking these three factors together into account, we define the value range for each data to estimate uncertainties. The specific explanations are as follows.
\begin{enumerate}
\item  The most significant uncertainty comes from the RMS noise of the temperature. As in the discussion in Section \ref{sec:radex}, our LVG code requires input \CO\ \Jone\ and \CO\ \Jthree\ temperatures. 
If we consider $T_\textrm{c}$(\Jone)$-$RMS and $T_\textrm{c}$(\Jthree)$+$RMS, the code outputs the smaller \N(CO) and the larger $n$(\HTWO), compared to the result of $T_\textrm{c}$(\Jone) and $T_\textrm{c}$(\Jthree). When we consider instead $T_\textrm{c}$(\Jone)$+$RMS and $T_\textrm{c}$(\Jthree)$-$RMS, the code provides larger \N(CO) and smaller $n$(\HTWO). We define the computed \N(CO) and $n$(\HTWO) in this way as the upper and lower limits for the range of data values, respectively.
\item  The calculation of \N(CO) requires the line width $\delta v$. We take the arithmetic mean of $\delta v$(\Jone) and $\delta v$(\Jthree) 
in our calculation. Here we put the larger $\delta v$ of both $T_\textrm{c}$(\Jone) and $T_\textrm{c}$(\Jthree) in the calculation to get the upper limit of the \N(CO) range. We take the smaller one of the two individual $\delta v$ to calculate the lower limit of the \N(CO) range.
\item  The \tk\ assumption is also relevant to the calculation results. According to Figure \ref{fig:tk}, \tk$=$15 K would lead to a large $n$(\HTWO). Therefore, we choose 30$\pm $10 K as a reasonable \tk\ range.
\end{enumerate}
We have run the LVG code with these $T_\textrm{c}\pm$RMS noise, $\delta v$, and \tk\ simultaneously, and obtain a range of data for \N(CO) and $n$(\HTWO). For each group of data, we define the median value as the result. The upper and lower limits of the range for this median pixel are found. The upper and lower uncertainties for each data set are the differences between the upper and lower limits of the range and the median value, respectively. The median values of $n$(\HTWO) and \N(CO) and their uncertainties are summarized in Table \ref{table:nco}.

\subsection{Comparison of \N(CO) under the two methods }\label{sec:3methods}

\begin{figure}
    \centering
	\includegraphics[width=0.6\columnwidth]{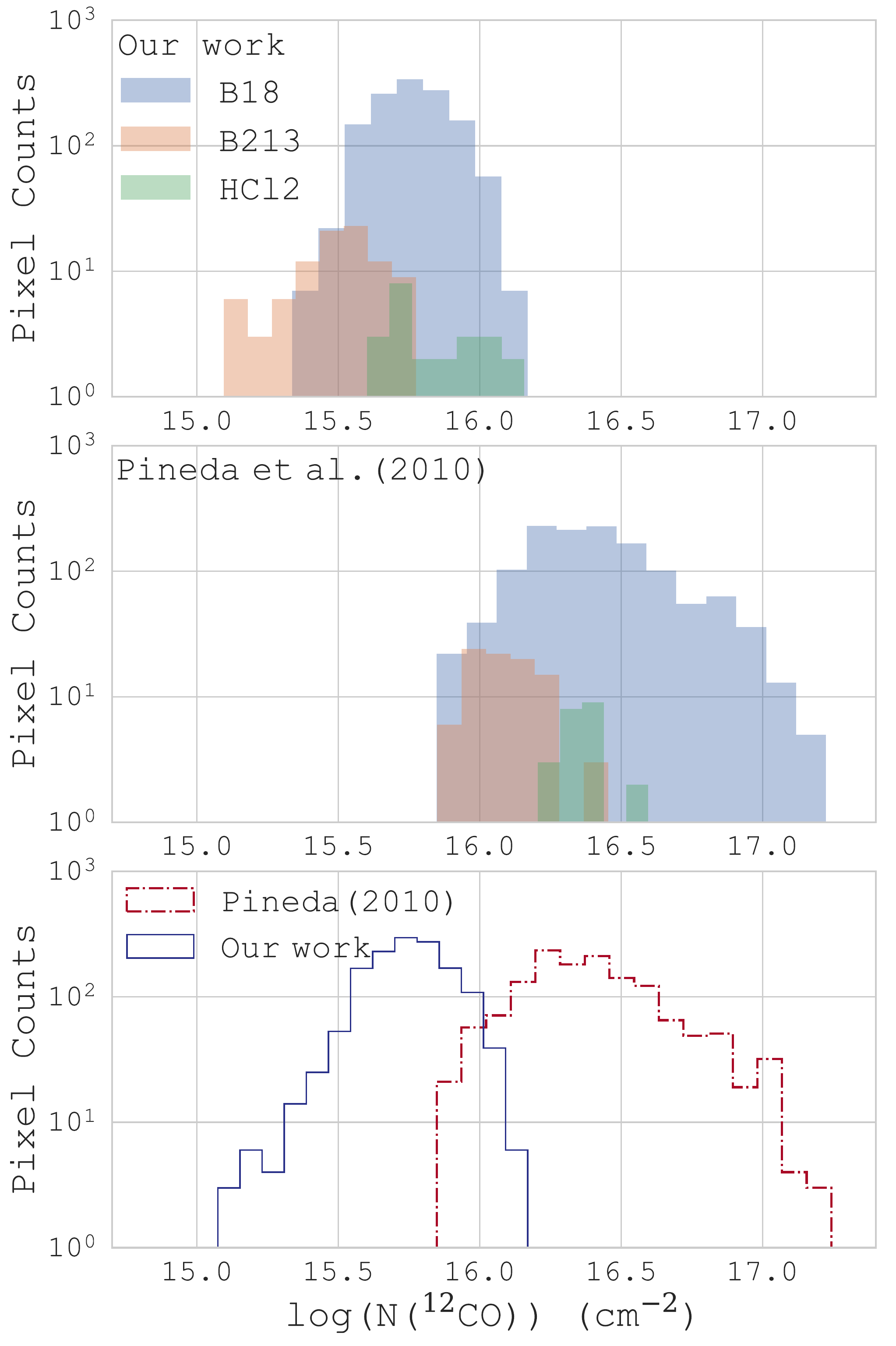}
    \caption{Histograms of \CO\ column density \N(CO) from our calculation by LVG code (top) and the result from \citet{2010ApJ...721..686P} by RADEX program (middle) for the B18 (blue), B213-L495 (orange), HCl2 (green) clouds within Taurus mask 1 region. And comparison of the our result (blue solid line) and \citep{2010ApJ...721..686P} (red dashdot line) methods of the three clouds with a total of 1395 pixels (bottom).}
    \label{fig:total_hist}
\end{figure}

The free parameters in \citet{2010ApJ...721..686P} are  $n$(\HTWO), $N(^{12}\textrm{CO})/\delta v$ and the \CO$/$\CCO\ abundance ratio. The observable parameters are \CO\ \Jone\ and \CCO\ \Jone\ intensities. 
The free parameters in our study are $n$(\HTWO) and $N(^{12}\textrm{CO})/\delta v$. The observable parameters are \CO\ \Jone\ and \CO\ \Jthree\ temperatures. 
Observations of the \Jone\ and \Jthree\ lines can output a single group of the best fitted $n$(\HTWO) and $N(^{12}\textrm{CO})/\delta v$ within the inverted grid of models.

We compare the histograms of \N(CO)\ for the three clouds B18, B213-L1495, HCl2, and the general results in Figure \ref{fig:total_hist}. In Table \ref{table:nco}, we summarize the \N(CO)\ and $n$(\HTWO) from our data, the \N(CO)\ from \citet{2010ApJ...721..686P}, and the correction ratio between the two sets. We compare the following aspects for the results of these two studies in what follows.
\begin{enumerate}
\item  Overall, \N(CO) calculated presently is $5.7^{+1.8}_{-0.4} \times 10^{15}\ \textrm{cm}^{-2}$, which is 0.24 times of the results from \citet{2010ApJ...721..686P}, $2.35^{+0.58}_{-1.53} \times 10^{16}\ \textrm{cm}^{-2}$. In the diffuse portion of the molecular clouds, the \N(CO) for mask 1 from either work is comparable.

\item  The two studies assume different values of \tk. The assumption of \citet{2010ApJ...721..686P}, $T_\textrm{k}=15$ K, is not satisfactory here as it would result in $n$(\HTWO) of $10^4\ \textrm{cm}^{-3}$, which is far too high for regions at the edge of clouds which are considered to be not dense. Our result of $n$(\HTWO)$=3.3^{+7.0}_{-1.8} \times 10^{3}\ \textrm{cm}^{-3}$ at \tk$=$30 K is somewhat higher than \citet{2010ApJ...721..686P}, but is not unreasonable. A similar finding at the edge of the dark cloud B5 has been published by \citet{1982ApJ...261..513Y}.

\item   The measurement of the thermal pressure ($\left<P_\textrm{th}/k\right>=nT$) of the gas from \citet{2010ApJ...721..686P} is between $4\times10^3$ to $1\times10^4\ \textrm{K}\ \textrm{cm}^{-3}$ (with a $n$(\HTWO) range of 250--700 $\textrm{cm}^{-3}$). Our value of $\left<P_\textrm{th}/k\right>=10^5\ \textrm{K}\ \textrm{cm}^{-3}$ is still within a reasonable range of the observed thermal pressure for molecular clouds in the Galactic plane \citep{1993AIPC..278..311S}. 
\end{enumerate}

Among the three clouds in Taurus, B18 includes a large amount of data, with a good sensitivity. Both of the calculation methods indicate that the \N(CO) in the B213 cloud is lower than in the other two regions. Because the limited number of selected pixels may be located where the gas is more diffuse. It does not represent the case of the entire B213 cloud.

\section{Discussion}\label{sec:nh2}
\begin{figure}
    \centering
	\includegraphics[width=0.6\columnwidth]{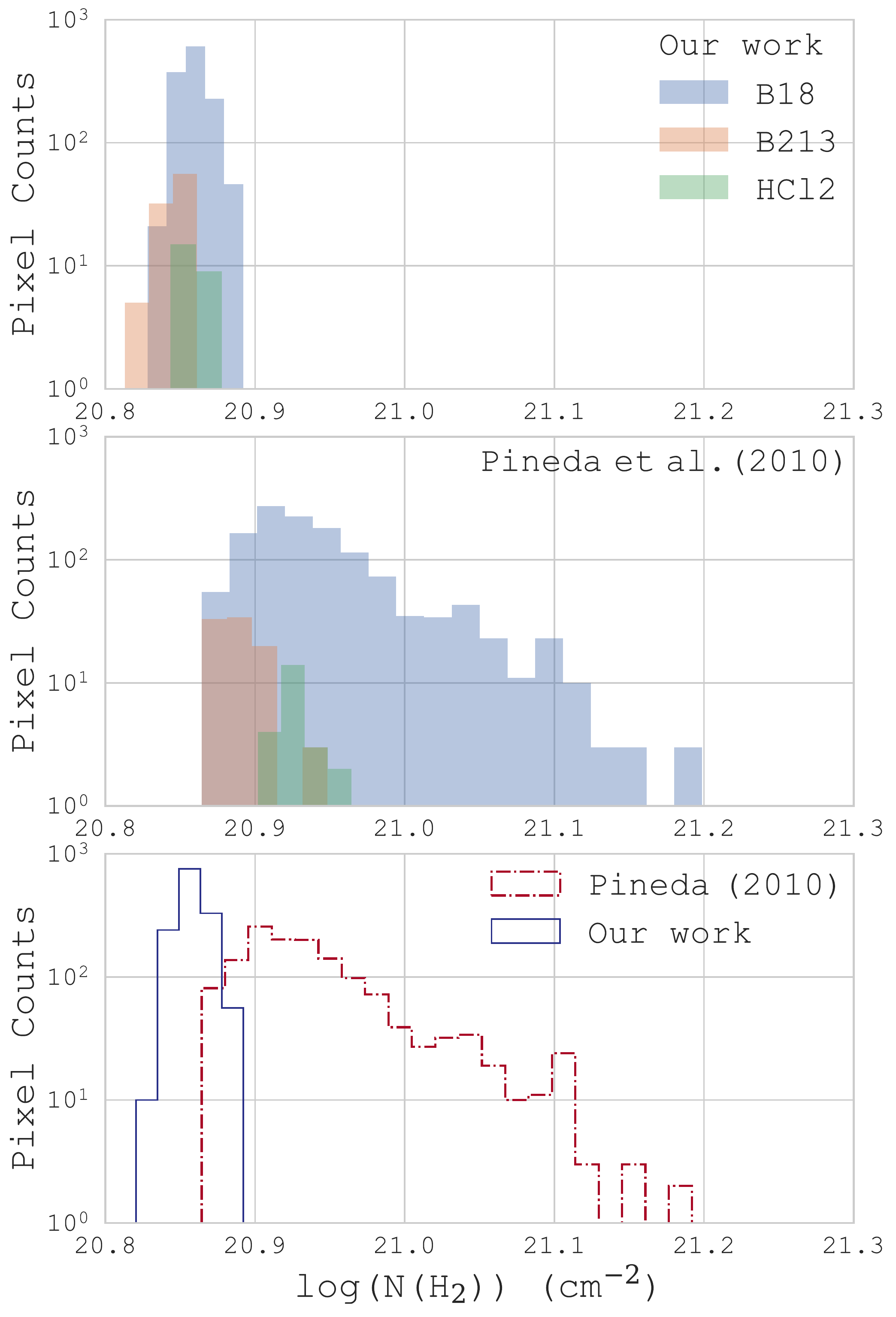}
    \caption{Histograms of \N(\HTWO) from our calculation with LVG code (top panel) and the result of \citet{2010ApJ...721..686P} by RADEX program (middle panel) for the B18 (blue), B213-L495 (orange), HCl2 (green) clouds within Taurus mask 1 region. And comparison of the \N(\HTWO) from our (blue solid line) and \citet{2010ApJ...721..686P} (red dashdot line) results for the three clouds with a total of 1395 pixels (bottom panel).}
    \label{fig:nh2}
\end{figure}

For low column density regions such as mask 1 and mask 0, fractional abundance of carbon monoxide, \abundance, may vary with CO column density over a wide range. For mask 1, \citet{2010ApJ...721..686P} used 2MASS extinction to convert to \N(\HTWO), assuming $N(\textrm{H}_2)/A_\textrm{V}\, =\, 9.4\times 10^{20}\, \textrm{cm}^{-2}\, \textrm{mag}^{-1}$ \citep{1978ApJ...224..132B}, and fitted a relation between \N(\HTWO) and \N(CO). We apply the same \N(CO)--\N(\HTWO) relation from \citet{2010ApJ...721..686P}. For every pixel, we have derived \N(\HTWO) corresponding to all \N(CO) data according to the equation, log(\N(\HTWO))= 0.03887$\times$log(\N(CO))$^3$ - 1.779$\rm \times$log(\N(CO))$^2$ +27.175$\times$log(\N(CO)) -117.71, from Figure 14 of \citet{2010ApJ...721..686P}. 
Uncertainties of \N(\HTWO) come from the \N(CO) range of values. We input the upper and lower limits of the \N(CO) range into this equation to calculate the upper and lower limits of \N(\HTWO) range. We summarize the median $\rm \textit{N}(H_2)$ data and their uncertainties for both studies in Figure \ref{fig:nh2} and Table \ref{table:nco}.

\begin{table}[]
\caption{Physical parameters in the Mask1 with the CO \Jthree\, and \Jone\, regions}
\label{table:nco}
\centering
\renewcommand{\arraystretch}{1.45}
\setlength{\tabcolsep}{1.6mm}{
\begin{tabular}{@{}llllllllll@{}}
\toprule
Region             & Pixel  & Paper     & $n$(\HTWO)        & \N(CO)                 & \N(\HTWO)                        \\
                   & Number & for \N(CO)&($1\times 10^3$ cm$^{-3}$)  &($1\times 10^{15}$ cm$^{-2}$)          & ($1\times 10^{20}$ cm$^{-2}$) \\ \midrule
B18                & 1276   & our      & 3.3$^{+8.9}_{-2.2}$ & 5.8$^{+2.1}_{-1.2}$         & 7.2$^{+0.2}_{-0.1}$  \\ 
                   &        & Pineda    &              & 24.6$^{+14.9}_{-11.7}$ (24\%) & 8.6$^{+0.9}_{-0.8}$ (84\%)           \\
HCl2               & 25     & our       & 7.8$^{+34.9}_{-4.2}$      & 6.0$^{+2.0}_{-1.2}$          & 7.2$^{+0.2}_{-0.1}$                   \\
                   &        & Pineda    &      & 23.4$^{+21.4}_{-3.6}$ (26\%)   & 8.5$^{+1.3}_{-0.2}$ (85\%)            \\
B213-L1495         & 94     & our       & 3.2$^{+4.2}_{-1.3}$  & 3.2$^{+1.1}_{-0.8}$          & 7.0$^{+0.1}_{-0.07}$                \\
                   &        & Pineda    &     & 11.9$^{+7.3}_{-4.7}$ (27\%)   & 7.7$^{+0.5}_{-0.4}$ (91\%)            \\
Above clouds & 1395   & our       & 3.3$^{+7.0}_{-1.8}$ & 5.7$^{+1.8}_{-0.4}$   & 7.2$^{+0.1}_{-0.04}$                  \\
                   &        & Pineda    &              & 23.5$^{+5.8}_{-15.3}$ (24\%)  & 8.5$^{+0.4}_{-1.1}$ (85\%)           \\ \bottomrule
                   
\end{tabular}
}
\footnotesize
\begin{flushleft}
	\textbf{Note}. We defined the correction of our result to the result of \citep{2010ApJ...721..686P} as the percentage given in parentheses.
\end{flushleft}
\end{table}

In our calculation, the median CO-derived \N(\HTWO) is $7.2^{+0.1}_{-0.04}\times10^{20}\ \textrm{cm}^{-2}$, which is 85\% of \citep{2010ApJ...721..686P} results. The \N(\HTWO) results are not much different between the two studies, even though our \N(CO) is 24\% of \citet{2010ApJ...721..686P} \N(CO). These \N(CO) data cannot accurately measure the changes in \N(\HTWO). This is because the \N(\HTWO)--\N(CO)
conversion is insensitive in the range of $\sim10^{15}\ \textrm{cm}^{-2}$--$10^{16}\ \textrm{cm}^{-2}$. When \N(CO) decreases 10 times, \N(\HTWO) decreases only 0.9 times. Here we conclude that there is almost no change in CO-derived \N(\HTWO) compared to \citet{2010ApJ...721..686P}.

However, there is a large difference in the \HTWO\ density and kinetic temperature between the two studies. While \citet{2010ApJ...721..686P} find densities around 500 $\textrm{cm}^{-3}$ for a temperature of 15 K, we find densities 6 times higher and a temperature $\sim$2 times higher. The \Jthree\ \CO\ observations bring important constraints which questions the previous study. However, our observations are close to the edges of the clouds and may not be representative of the mask 1 region (no \CCO\ detected) in general.

\section{Conclusions}\label{sec:Conclusions}
In the Taurus molecular cloud, we estimated the \N(CO), $n$(\HTWO), and \N(\HTWO) in the non-\CCO\, detection region using JCMT \CO\,\Jthree\, and FCRAO \CO\,\Jone\, survey data \citep{2010MNRAS.405..759D,2008ApJS..177..341N,2008ApJ...680..428G}. Our measurements include parts of the edges of the B18, HCl2, and B213-L1495 clouds, containing a total of 1395 pixels. We draw the following conclusions: 
\begin{enumerate}
\item  In mask 1, we have run a LVG code with \CO\,\Jone\, and \Jthree\ data to calculate a \N(CO) of $5.7^{+1.8}_{-0.4}\times 10^{15}\ \textrm{cm}^{-2}$, about 24\% of \citet{2010ApJ...721..686P} finding. 
And we estimated $n$(\HTWO) to be $3.3^{+7.0}_{-1.8} \times 10^{3}$ $\textrm{cm}^{-3}$ under the assumption of \tk$=$30 K.

\item  We have estimated the \N(H$_2$) pixel by pixel, using the \N(CO)--\N(\HTWO) relation \citep{2010ApJ...721..686P}. The derived \N(\HTWO) almost did not change from the \citep{2010ApJ...721..686P}'s result. The median \N(H$_2$) is $7.2^{+0.1}_{-0.04}\times10^{20}\ \textrm{cm}^{-2}$.

\end{enumerate}

Overall, the calculation of \N(CO), $n$(\HTWO), and the assumption of \tk\ in mask 1 in \citet{2010ApJ...721..686P} are different from ours. Using two transitions of \CO\ data, we measured a lower \N(CO) and a higher $n$(\HTWO), assuming a higher \tk. This measurement of only 1395 pixels is suggestive for future studies of the physical conditions of cloud edges for dark clouds like Taurus. More sky area coverage and more systematic measurements are needed.

\section*{Acknowledgements}
This work is supported by the National Natural Science Foundation of China (NSFC, Grant  No. 11988101, No. 11725313, No. U1931117) and the International Partnership Program of Chinese Academy of Sciences (Grant No. 114A11KYSB20210010). This research was carried out in part at the Jet Propulsion Laboratory, which is operated by the California Institute of Technology under a contract with the National Aeronautics and Space Administration (80NM0018D0004). C.W. is supported by the Natural Science Foundation of Jiangsu Province(Grants No BK20201108).


\bibliographystyle{raa}



\end{document}